\documentclass[aip,graphicx]{revtex4-1}

\usepackage{graphicx}% Include figure files
\usepackage{dcolumn}% Align table columns on decimal point
\usepackage{bm}% bold math
\usepackage[utf8]{inputenc}
\usepackage[T1]{fontenc}
\usepackage{mathptmx}
\usepackage{etoolbox}

\draft % marks overfull lines with a black rule on the right

\begin{document}

%\preprint{}

\title{Lineshapes in Pump-Probe Spectroscopy of Polaritons} %Title of paper



\author{Luca Nils Philipp}
\affiliation{Institut für Physikalische und Theoretische Chemie, Universität Würzburg, Emil-Fischer Straße 42, 97074 Würzburg, Germany}
\author{Eva Münzel}
\affiliation{Institut für Physikalische und Theoretische Chemie, Universität Würzburg, Emil-Fischer Straße 42, 97074 Würzburg, Germany}
\author{Julian Lüttig}
\affiliation{Department of Physics, University of Michigan, 450 Church Street, Ann Arbor, Michigan 48109, USA}
\affiliation{Institut für Physikalische und Theoretische Chemie, Universität Würzburg, Am Hubland, 97074 Würzburg, Germany}
\author{Roland Mitri\'c}
\email{roland.mitric@uni-wuerzburg.de}
\affiliation{Institut für Physikalische und Theoretische Chemie, Universität Würzburg, Emil-Fischer Straße 42, 97074 Würzburg, Germany}


\date{\today}

\begin{abstract}
Forming new hybrid quasiparticles by strong light-matter coupling is a promising tool for tailoring photophysics and photochemistry of molecules. Thus, the ultrafast dynamics of polaritons formed upon strong light-matter coupling has been extensively studied by pump-probe spectroscopy. Although it was predicted that the partial photonic character of polaritons should shorten their lifetime compared to purely molecular excited states, many studies do not observe this effect. So far, the unexpected longevity of the spectral signatures was either explained by relaxation into a manifold of so-called dark states or by other uncontrolled effects that change the properties of cavity materials. In order to resolve these issues, we investigate here the dependence of the lineshape of pump-probe spectra of polaritons on the ratio of photonic and molecular character. 
Furthermore, by phenomenologically including relaxation to dark states, we find that it is possible to spectrally resolve this relaxation process by observing a characteristic phase flip in the pump-probe signal. Our results show that the signatures of various effects and their contributions to the polariton dynamics can be disentangled from the spectral lineshapes.
\end{abstract}


\maketitle %\maketitle must follow title, authors, abstract and \pacs

\section{Introduction}
Molecular polaritons are hybrid quasiparticles that arise when confined electromagnetic fields interact strongly with the excited states of molecules.\cite{fassioli_femtosecond_2021} The polaritonic states inherit both photonic and molecular properties. In molecular systems, polaritons are usually formed by coupling molecular vibrations or electronic excitations to strong electromagnetic fields created by optical microcavities or plasmonic systems. For example, it has been shown that vibrational polariton formation can alter the potential energy surfaces of molecules such that chemical reaction rates are modified\cite{thomas_tilting_2019, thomas_ground-state_2016, ahn_modification_2023, galego_cavity-induced_2015, hutchison_modifying_2012, sau_modifying_2021, schwartz_reversible_2011, lather_cavity_2019}. In this work, we will focus on polaritons, which arise by coupling electronically excited molecular states to the electromagnetic field modes. Under electronic strong-coupling conditions, theoretical studies suggest an enhancement of the exciton transport rate in organic semiconductors\cite{feist_extraordinary_2015, schachenmayer_cavity-enhanced_2015, gonzalez-ballestero_harvesting_2015}. However, experimental observations of this effect so far are ambigious and the role of different polariton states in enhancing transport properties remains unclear\cite{zhong_energy_2017, zhong_non-radiative_2016, nagarajan_conductivity_2020, orgiu_conductivity_2015, hou_ultralong-range_2020, rozenman_long-range_2018}. Furthermore, it is still an open question how polaritons interact with each other and with other states that are not involved in the polariton formation process\cite{fassioli_femtosecond_2021}.

To answer these questions, which are related to the excited state dynamics of molecular polaritons, nonlinear spectroscopy techniques have been employed, in particular transient pump-probe spectroscopy (PP). For example, Virigil and coworkers have used PP spectroscopy to study femtosecond relaxation processes in an organic-semiconductor microcavity.\cite{virgili_ultrafast_2011} In addition to further studies on the relaxation processes\cite{schwartz_polariton_2013, chen_tracking_2025}, other authors also investigated different nonlinear effects of polaritons like Rabi-contraction\cite{delpo_polariton_2020}, motional narrowing\cite{wanasinghe_motional_2024, ying_theory_2024}, and quasi-particle annihilation\cite{wu_efficient_2025, buttner_probing_2025}.

An unsolved problem for many PP spectroscopic studies of polaritons is that the spectral features often have strikingly longer lifetimes than would be theoretically expected from the partially photonic character of polaritons.\cite{schwartz_polariton_2013, delpo_polariton_2020, renken_untargeted_2021, wu_optical_2022} This led Renken et al. to investigate the long-lived features in their PP spectra of a microcavity containing an organic dye.\cite{renken_untargeted_2021} They found that the features can be explained by a combination of ground state bleach of the organic dye, small changes in the refractive indices of the cavity materials and organic dyes, as well as a change in the thickness of the organic layer. 

An alternative explanation for the longevity of the spectral signals in PP spectra of polaritons is a fast relaxation from the polariton states to a manifold of so-called dark states (DS).\cite{gonzalez-ballestero_uncoupled_2016, botzung_dark_2020, schwartz_polariton_2013, georgiou_generation_2018, xiang_state-selective_2019} Coupling of $N$ molecules to a single electromagnetic field mode always leads to the formation of $N-1$ DS, whereas only two polariton states are formed. It has been argued that the DS can act as a sink for excitations due to their large number.\cite{scholes_entropy_2020} Compared to the polariton states, the DS are linear combinations of purely molecular states and do not posses any photonic character. Therefore, the DS are not effected by lifetime shortening as expected for the polariton states.

Renken et al. showed that the lineshape of each contribution of the above mentioned effects has a distinct dependence on the incidence angle of the exciting laser pulses on the sample.\cite{renken_untargeted_2021} By varying the incidence angle, cavity field modes with different resonance frequencies are coupled to the molecules such that polaritons with different ratios of molecular and photonic character are excited.\cite{rodel_anisotropic_2025, buttner_probing_2025} How the incidence angle is connected with the resonance energy of the cavity strongly depends on the type of cavity. Comparing the dependence of the nonlinear response of polaritons and the untargeted effects on the incidence angle, would allow one to distinguish between these two. However, it is so far not clarified how the lineshape of the "true" polariton spectral features depend on the incidence angle. Furthermore, there is no way to unambiguously reveal if the DS are being populated.

In this work, we address this issue by investigating the influence of the cavity resonance energy and DS relaxation on the lineshape of PP spectra of polaritons. First, we simulate the dependence of the lineshape of PP spectra of polaritons without DS relaxation on the cavity resonance energy. Subsequently, we phenomenologically include relaxation from the polariton states to the DS manifold to investigate if this process effects the dependence of the lineshape on the cavity resonance energy. The simulations are carried out within the response function formalism, which assumes a perturbative treatment of the interaction between the system and the exciting laser pulses. For its calculation, eigenstates of the polaritonic system are needed, which we determine within the Tavis-Cummings (TC) model\cite{tavis_exact_1968, tavis_approximate_1969}.

\section{Theoretical Methods}
\subsection{Tavis-Cummings Model}
The TC model describes a system consisting of $N$ two-level systems interacting with a single quantized electromagnetic field mode. The Hamiltonian of the TC model is given by

\begin{equation}
    H_\mathrm{TC} = \hbar\omega_m\sum_{i=1}^N\sigma_i^+\sigma_i^- + \hbar\omega_ca^\dag a + \hbar g \sum_{i=1}^N\left(a^\dag\sigma_i^- + \sigma_i^+a\right),
\end{equation}

with the light-matter coupling constant $g$, the annihilation (creation) operator of the bosonic field mode $a$ ($a^\dag$) of frequency $\omega_c$, and the Pauli raising (lowering) operator of the \textit{i}-th two-level system $\sigma_i^-$  ($\sigma_i^+$) of frequency $\omega_m$. Since the Hamiltonian commutes with the generalized number operator $n = a^\dag a + \sum_{i=1}^N\sigma_i^+\sigma_i^-$, its eigenstates can be divided into subspaces with different numbers of $n$ quanta distributed along the two-level systems and the cavity mode. The eigenstates of the single-excitation subspace are sufficient to describe the linear response of the system. However, if non-linear properties are investigated, eigenstates with higher excitation numbers become important. In the case of third-order processes involved in PP spectroscopy, one needs single- and two-particle eigenstates to calculate the response of the system. Analytical solutions of the single- and two-particle eigenstates are well known in the resonant case.

Usually, the eigenstates of the TC model are expressed in the so-called bare basis. Within the bare basis, a state is defined by the tensor product of eigenstates of the isolated $N$ two-level systems and the electromagnetic field mode $|\psi\rangle\otimes| \phi\rangle$. Here, $|\psi\rangle$ is the state of the isolated two-level systems and $|\phi\rangle$ is the state of the field mode. While $|\phi\rangle$ is always a number state $|n\rangle$, the two-level systems are either all in the ground state $|g\rangle$, one two-level system is excited $|e_i\rangle$ or two two-level systems are excited $|e_ie_j\rangle$ ($i\neq j$). Then all basis states with up to two excitations distributed along the system are contained in $\{|g\rangle\otimes|n\rangle, |e_i\rangle\otimes|n\rangle, |e_ie_j\rangle\otimes|n\rangle\}$. By diagonalizing the TC Hamiltonian in this basis mixed polaritonic eigenstates are obtained from which the transition amplitudes can be calculated.

In their works on organic semiconductor microcavities, Lidzey et al. have introduced the concept of the visibility of polaritonic states, which is a measure for the linear absorption strength of polariton states.\cite{lidzey_experimental_2002, lidzey_photon-mediated_2000, lidzey_room_1999} They define the visiblity $I$ by $I=|\alpha|^2$, where $\alpha$ is the coefficient of the photonic basis state in the basis expansion of the full polariton state. While this definition only applies to transitions between the GS and single-particle eigenstates of the TC model, it is readily generalized by replacing the coefficient $\alpha$ with the matrix elements of the hermitian operator $a+a^\dagger$ representing the electric field. In this way, the visibility of a transition between two eigenstates of the TC model $|\phi\rangle$ and $|\psi\rangle$ is defined as $I_{\phi\rightarrow\psi}=|\langle\phi|a+a^\dagger|\psi\rangle|^2$. In the following, we will refer to the matrix elements of $a+a^\dagger$ as photonic transition moments.

\begin{figure}
  \includegraphics[]{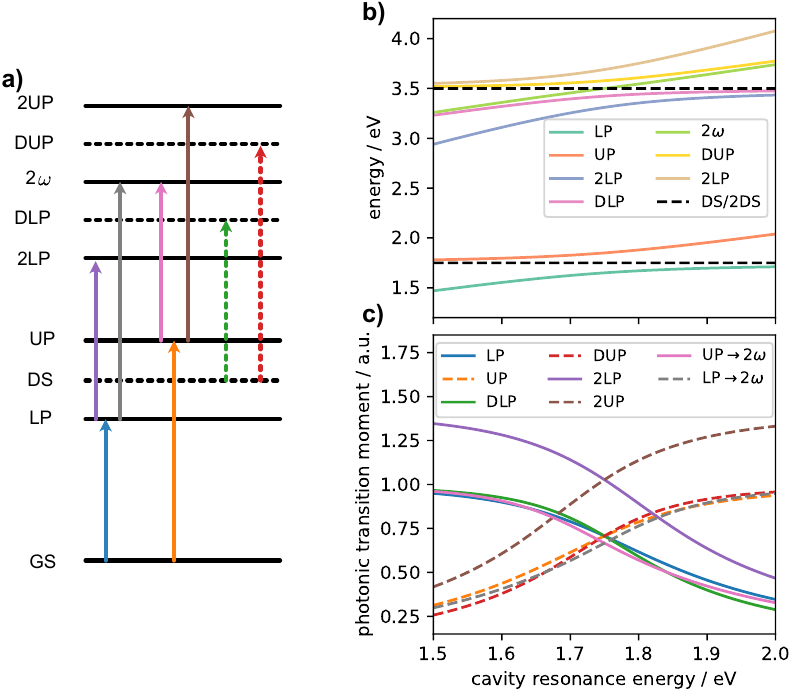}
  \caption{\label{figure1}a) Energy level structure of the one- and two-particle eigenstates of the TC model. Polariton states without DS character are indicated by horizontal lines, while those with DS character are indicated by dashed horizontal lines. Allowed transitions between polariton states are indicated by arrows. Dotted arrows are used for transitions between states with DS character. The color of the arrows refers to the color of the lines of the corresponding transition amplitude in Fig. \ref{figure1}c. b) Energies of the one- and two-particle eigenstates as a function of the cavity resonance energy. c) Photonic transition moments between the eigenstates of the TC model. LP-(UP-)like transitions are indicated by solid (dashed) lines. States, which can only be reached by one transition, are labeled with the final state of the transition.}
\end{figure}

The eigenstates of the TC model without detuning and the transition moments will be described in the following. A schematic representation of the energy levels of the TC model is shown in Fig. \ref{figure1}a. The single-particle eigenstates of the TC Hamiltonian consist of the lower polariton (LP), upper polariton (UP) and $N-1$ DS. The wavefunctions of the UP and LP can be expressed as

\begin{equation}
    |\mathrm{UP}/\mathrm{LP}\rangle = \frac{1}{\sqrt{2}} |g\rangle\otimes|1\rangle \pm \sum_{i=1}^N |e_i\rangle\otimes|0\rangle,
\end{equation}

with eigenenergies $E_{\mathrm{UP/LP}}=\hbar\omega \pm \hbar g\sqrt{N}$. LP as well as UP can be excited from the ground state with corresponding photonic transition moment $\langle g|\otimes\langle0|a+a^\dagger|\mathrm{UP/LP}\rangle=\frac{1}{\sqrt{2}}$.
All dark states are linear combinations of bare molecular states without excitations of the electromagnetic field mode

\begin{equation}
    |\mathrm{DS}\rangle=\sum_{i=1}^N c_i |e_i\rangle\otimes|0\rangle.
\end{equation}

Since the coefficients must fulfill $\sum_{i=1}^N c_i=0$, there is no contribution from excited photon modes in the wavefunction of the DS. Thus, they have vanishing photonic transition moment and cannot get directly excited in linear absorption spectroscopy. The DS have an energy of $E_\mathrm{DS}=\hbar\omega$.

Generally, the two-particle eigenstates can be thought of as two simultaneous excitation of single-particle eigenstates. Since there are two different kinds of states in the single-particle manifold, polariton states (UP and LP) and DS, there are three different kinds of states in the two-particle manifold, polariton-polariton states, polariton-DS states and DS-DS states. As in the case of transitions between the ground states and single-particle states, transitions between single- and two-particle eigenstates are generally only possible if polariton character is changed during the process. This means that only transitions between polariton states and polariton-polariton states and transitions between DS and polariton-DS are allowed. 

Polariton-polariton states consist of two simultaneous excitations of either LP or UP and are called second lower polariton (2LP), second upper polariton (2UP), and $2\omega$ state. While the 2LP (2UP) consists of two simultaneous excitations of the LP (UP), the $2\omega$ state is the simultaneous excitation of LP and UP. The 2LP and 2UP states are the energetic lower and upper bounds of the two-particle eigenstates with eigenenergies $E_{\mathrm{2UP/2LP}}=2\hbar\omega\pm2\hbar g\sqrt{N-\frac{1}{2}}$. Their wavefunction can be written as

\begin{equation}
        |\mathrm{2UP/2LP}\rangle=\frac{1}{\sqrt{2}} \sqrt{\frac{N}{(2N-1)}}|g\rangle\otimes|2\rangle  \pm\frac{1}{\sqrt{2N}} \sum_{i=1}^N |e_i\rangle\otimes|1\rangle +\frac{1}{\sqrt{N(2N-1)}}\sum_{i=1}^N\sum_{j>i}^N|e_i e_j\rangle\otimes|0\rangle.
\end{equation}

2UP (2LP) can only be excited from the UP (LP) with photonic transition moment $\langle \mathrm{UP}|a+a^\dagger |\mathrm{2UP}\rangle=\langle \mathrm{LP}|a+a^\dagger |\mathrm{2LP}\rangle=\frac{1}{2}+\sqrt{\frac{N}{2(2N-1)}}$.
The $2\omega$ state can also get excited from LP and UP. The corresponding wavefunction is given by

\begin{equation}
        |2\omega\rangle= \sqrt{\frac{N-1}{2N-1}} |g\rangle\otimes|2\rangle -\sqrt{\frac{2}{(N-1)(2N-1)}} \sum_{i=1}^N \sum_{j>i}^N|e_i e_j\rangle\otimes|0\rangle,
\end{equation}

and the photonic transition moments are $\langle \mathrm{LP}|a+a^\dagger |2\omega\rangle=\langle \mathrm{UP}|a+a^\dagger |2\omega\rangle=\sqrt{\frac{N-1}{2N-1}}$. The eigenenergy of the $2\omega$ state is $E_{2\omega}=2\hbar\omega$.

Polariton-DS states are called dark lower (DLP) and dark upper polaritons (DUP). Since these states are simultaneous excitations of LP or UP and a DS, there are each $N-1$ DLPs and DUPs within the manifold of two-particle states. The DUPs (DLPs) are energetically located between the 2UP (2LP) and the $2\omega$ state at an energy of $E_{\mathrm{DUP/DLP}}=2\hbar\omega\pm\hbar g\sqrt{N-2}$. Their wavefunction can be expressed as

\begin{equation}
        |\mathrm{DUP/DLP}\rangle=\sum_{i=1}^N c_i |e_i\rangle\otimes|1\rangle \pm \frac{1}{\sqrt{N-2}} \sum_{i=1}^N\sum_{j>i}^N(c_i+c_j )|e_i e_j\rangle\otimes|0\rangle.
\end{equation}

Since the DS and DUPs (DLPs) are parametrized with the same set of coefficients $c_i$, DUPs and DLPs can only be excited from one of the DS with photonic transition moment $\langle \mathrm{DS}|a+a^\dagger |\mathrm{DLP}\rangle=\langle \mathrm{DS}|a+a^\dagger |\mathrm{DUP}\rangle=\frac{1}{\sqrt{2}}$. Finally, there is a manifold of DS-DS states (2DS), which are simultaneous excitations of two DS with an energy of $E_{2DS}=2\hbar\omega$. The corresponding wavefunctions are given by 

\begin{equation}
    |\mathrm{2DS}\rangle=\sum_{i=1}^N \sum_{j>i}^N c_{ij} |e_i e_j\rangle\otimes|0\rangle,
\end{equation}

where the coefficients must satisfy $\sum_{j\neq i}^N c_{ij}=0$. These states cannot get excited from any of the one-particle states, since they do not have any contribution from excited photon states.

The analytic results presented above build a good intuition, how the state structure of the TC model looks like. However, if detuning is introduced, there are no analytical solutions available. Thus, we numerically diagonalize the Hamiltonian of the TC model with parameters $N=5$, $\hbar g\sqrt{N}=0.1\ \mathrm{eV}$, and $\hbar\omega_m=1.75\ \mathrm{eV}$ and variable resonance energies of the cavity mode to obtain the eigenstates and the photonic transition moments depending on the resonance energy of the cavity mode (Fig. \ref{figure1}b and \ref{figure1}c). The parameters are in a typical range for organic layers coupled to field modes, except for the number of coupled molecules, which is typically much larger.\cite{buttner_probing_2025} However, the difference in the size of photonic transition moments is only visible for a small number of molecules. From Fig. \ref{figure1}b it is evident that the single- and two-particle states are energetically well separated over the range of cavity resonance energies. At low cavity resonance energies, the energy of the UP is  close to that of the DS manifold due to its large molecular character, whereas the energy difference of LP and DS is large. Thus, the LP has only small molecular character, but larger photonic one. By increasing the cavity resonance energy, the LP slowly approaches the DS, while the energy difference between DS and UP increases. Meanwhile the molecular character of the LP and the photonic character of the UP also increase. If the transition energy of the two-level systems is in resonance with the cavity resonance energy, UP and LP have the same energetic difference to the DS and thus, have an equal share of photonic and molecular character. Further increasing the resonance energy of the cavity leads to a UP with mainly photonic character and a LP with mainly molecular character. 

Similar trends are observable for the two-particle eigenstates. The energy of the DLP and 2LP is well separated from the 2DS at low cavity resonance energies. Thus, they have mainly photonic character, while DUP and 2UP have mainly molecular character. Again, by increasing the resonance energy of the cavity, the energy of DLP and 2LP approach the 2DS. Furthermore, the energy difference between DUP (2UP) and the 2DS increases. Meanwhile, the molecular character of the DLP and 2LP increase, whereas the photonic character of DUP and 2UP increase. Generally the energy of the 2LP (2UP) is always lower (higher) than that of DLP (DUP). Since the $2\omega$ state represents the simultaneous excitation of LP and UP, its energy is approximately the mean value of 2LP and 2UP energies. For low cavity resonance energies, it is energetically close to the DLP, because it has similar character, as the UP has mainly molecular character here. For high cavity resonance energies, it is energetically close to the DUP energy as the LP has mainly matter character. 

Fig. \ref{figure1}c shows that there are mainly two different kinds of transitions between the polaritonic eigenstates. First, there are transitions, whose photonic transition moments increase with increasing cavity resonance energies. For the second kind of transitions, the photonic transition moments decrease with increasing cavity resonance energies. We will call the first (second) kind of transitions LP-(UP-)like transitions, since the GS to LP (UP) transition is among them. Besides the GS to LP transition, the DS to DLP, LP to 2LP, and UP to $2\omega$ transitions are all LP-like transitions. The size of the corresponding transition moments mainly follow the size of the photonic character of the LP. This holds likewise for the UP-like transitions, GS to UP, DS to DUP, UP to 2UP, and LP to $2\omega$, whose corresponding size of transition moments follow the size of the photonic character of the UP. Similar to the transition between the first and second excited states of the harmonic oscilator, the photonic transition moment of the LP to 2LP (UP to 2UP) transition is around $\sqrt{2}$ times larger than that of the other LP-(UP-)like transitions.

\subsection{Response Function Formalism}

To simulate third-order PP spectra, we utilize the response function formalism\cite{mukamel_principles_1995}, which will be outlined in the following. In general, the external laser field drives either the photonic or the matter degrees of freedom of the polaritonic system. For example, the matter degrees of freedom are driven by excitation through semitransparent mirrors in optical microcavities. However, if surface plasmon polaritons (SPP) are strongly coupled to molecules to create polaritonic states, the photonic degrees of freedom are driven by exciting the system in the Kretschmann geometry so that there is no direct interaction between the molecule and the external laser field.\cite{tame_single-photon_2008} In both cases the total Hamiltonian is given by

\begin{equation}
    H = H_\mathrm{TC} -\mu E(t),
\end{equation}

where the interaction with the external laser electric field $E(t)$ is treated semiclassically and $\mu$ is the corresponding transition moment. If the matter degrees are driven, the transition moment is the electronic transition dipole moment $\mu = \mu_m\sum_{i=1}^N \left(\sigma_i^++\sigma_i^-\right)$, where $\mu_m$ is the transition dipole moment of a single molecule. If the photonic degrees of freedom are driven, the transition moment is given by $\mu=\lambda(a+a^\dag)$ with the constant $\lambda$, which is connected to the coupling strength between the external electric field and the electromagnetic field mode of the cavity.\cite{tame_single-photon_2008} In the following, we will assume that the photonic degrees of freedom are driven, since this naturally leads to photonic transition moments to determine the transition strengths.

In a typical pump-probe experiment, the system interacts with two laser pulses. First, the pump pulse excites the system. Subsequently, the dynamics is probed by the probe pulse. To map the dynamics, the probe pulse is delayed by a delay time $T$ with respect to the pump pulse. After interaction with the two laser pulses, the system emits the polarization $P(t)$. The polarization depends on the systems density matrix $\rho(t)$ in the following way

\begin{equation}
    P(t) = \mathrm{Tr}\left(\rho(t)\mu \right).
\end{equation}

Within the response function formalism\cite{mukamel_principles_1995}, the systems density matrix is perturbatively expanded in powers of the external laser electric field

\begin{equation}
    \rho(t) = \sum_{n=0}^\infty \rho^{(n)}(t).
\end{equation}

The \textit{n}th order density matrix $\rho^{(n)}(t)$ is given by

\begin{eqnarray}
        \rho^{(n)}(t) = \left(-\frac{i}{\hbar}\right)^n \int_0^\infty dt_n\int_0^\infty dt_{n-1} \cdots \int_0^\infty dt_{1} 
        E(t-t_n)E(t-t_n-t_{n-1})\cdots E(t-t_n-\cdots-t_1) \\\nonumber
        U(t_1+\cdots+t_n)[\mu(t_1+\cdots+t_{n-1}), \left[\cdots\left[\mu(t_1),\left[\mu(0),\rho_{eq}\right]\right]\cdots\right]] U^\dagger(t_1+\cdots+t_n),
\end{eqnarray}

here $\rho_{eq}$ denotes the initial thermal density matrix, the time variables $t_1, \cdots, t_n$ denote the length of time intervals between successive interaction with the laser pulses, $U(t)=e^{-\frac{i}{\hbar}Ht}$ is the propagator of the unperturbed system and $\mu(t)=U(t)\mu U^\dagger(t)$ is the transition moment in the interaction picture. The perturbative expansion of the density matrix, induces the same kind of expansion for the polarization, such that the \textit{n}th order polarization is given by

\begin{equation}
        P^{(n)}(t) = \int_0^\infty dt_n\int_0^\infty dt_{n-1} \cdots \int_0^\infty dt_{1} 
        E(t-t_n)E(t-t_n-t_{n-1})\cdots E(t-t_n-\cdots-t_1) S^{(n)}(t_1, \cdots, t_n)
\end{equation}

with the \textit{n}th order response function defined as

\begin{equation}
        S^{(n)}(t_1, \cdots, t_n) = \left(-\frac{i}{\hbar}\right)^n \mathrm{Tr}(\mu(t_1+\cdots+t_n) 
    \left[\mu(t_1+\cdots+t_{n-1}), \left[\cdots\left[\mu(t_1),\left[\mu(0),\rho_{eq}\right]\right]\cdots\right]\right]).
\end{equation}

In the pump-probe geometry, the third-order polarization is the leading term in the perturbative expansion, since it is the first uneven nonlinear order, which includes interactions with pump and probe pulse. Therefore, PP spectra result predominantly from the detection of the emitted third-order polarization. After transformation from the interaction picture to the Schrödinger picture, the corresponding third-order response function is given by

\begin{equation}
    S^{(3)} (T, t)=\left(-\frac{i}{\hbar}\right)^3 \mathrm{Tr}\left(\mu U(t)[\mu,U(T)[\mu,[\mu, \rho_{\mathrm{eq}}]]]\right),
\end{equation}

where $T$ is the delay time between pump and probe pulse interactions. Under phase matching conditions, the generated third-order polarization is detected in the same direction as the probe pulse such that the sample has to interact twice with the pump pulse with wavevectors $\vec{k}_{pu}$ and $-\vec{k}_{pu}$. In this case, the third-order polarization is emitted in the direction of detection $\vec{k}_{pr} = \vec{k}_{pr} + \vec{k}_{pu} - \vec{k}_{pu}$. Since the two-interactions with the pump pulse happen simultaneously, we cannot assume strict time-ordering between the first two interactions. Thus, we have to consider both time orderings $+ \vec{k}_{pu} - \vec{k}_{pu}$ and $- \vec{k}_{pu} + \vec{k}_{pu}$. However, the interaction with the probe pulse is delayed by the delay time $T$ with respect to the pump pulse, i.e., we assume strict time-ordering between the pump and the probe pulse. We investigate the lineshape of the third-order PP spectra for short delay times such that at delay time $T=\tau$ no relaxation of the system has yet occurred, but the pump and the probe pulse do not longer overlap to form the so-called "coherent artifact".\cite{lorenc_artifacts_2002, mukamel_principles_1995} In this case, we can approximate the propagator at the specific delay time $T=\tau$ by $U(\tau) = 1$ such that the third-order response function simplifies to

\begin{equation}
    S^{(3)} (t)=\left(-\frac{i}{\hbar}\right)^3 \mathrm{Tr}\left(\mu U(t)[\mu,[\mu,[\mu, \rho_{\mathrm{eq}}]]]\right).
\end{equation}

\begin{figure*}
  \includegraphics[]{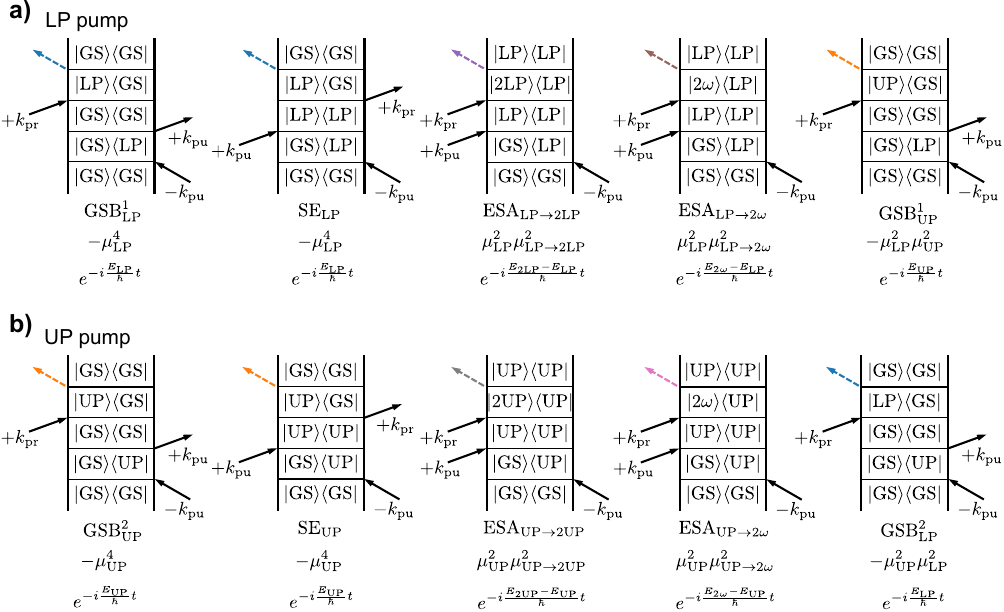}
  \caption{\label{figure2}a) (b))Double-sided Feynman diagrams for for excitation of LP (UP) with the pump. The color of the last dotted arrow of each Feynman diagrams indicates the energy at which the non-linear polarization is emitted. The amplitude and time-dependent phase factor accumulated during the time $t$ are written below every diagram.}
\end{figure*}

\begin{figure*}
  \includegraphics[]{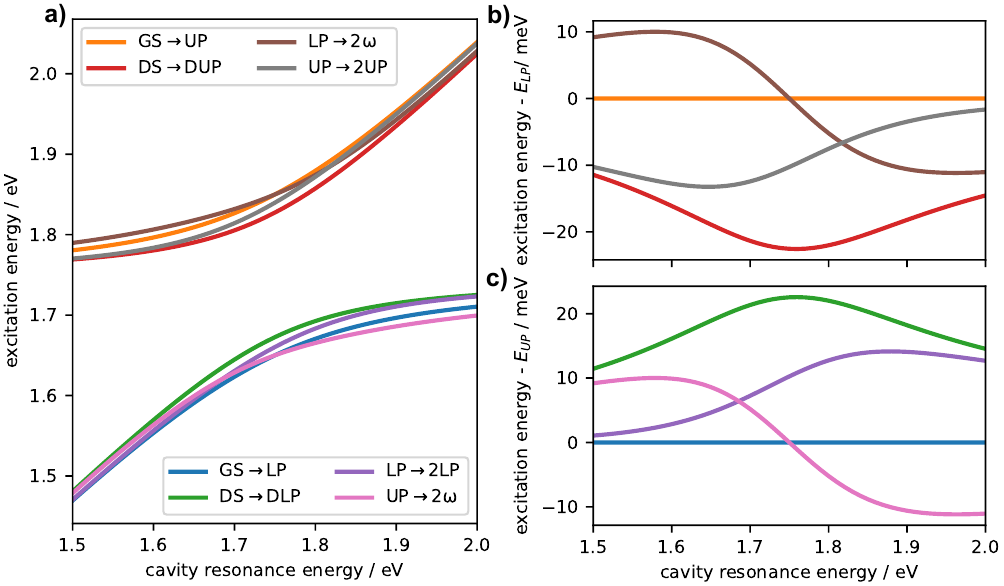}
  \caption{\label{figure3}a) Excitation energies of all allowed transitions between the eigenstates of the TC model depending on the resonance energy of the cavity. b) (c)) Excitation energies of UP-(LP-)like transition relative to the Excitation energy of the UP (LP).}
\end{figure*}

Since the first two interactions with the pump pulse happen simultaneously and we assume a short delay time, the response function only depends on the time $t$ between the interaction of the probe pulse with the system and the emission of the third-order polarization. The nested commutators in the response function lead to the emergence of various terms. To keep track of all possible terms, the Liouville space pathways are usually depicted by double-sided Feynman diagrams.\cite{mukamel_principles_1995}

The appearance of $n$ commutators in the response function leads to $2^n$ possible Liouville space pathways. However, by invoking the rotating wave approximation, time ordering and phase matching, the number of pathways can be reduced. Furthermore, we assume that that the pump pulse is either resonant with LP or UP, while in both cases LP and UP are probed such that only the Liouville space pathways in Fig. \ref{figure2} contribute to the third-order response. In PP spectrosocopy, there are generally three different kinds of Liouville space pathways, which differ in the photophysical processes they represent. First, there are ground state bleach (GSB) pathways, whose characteristic is that the system remains in the GS after the two interactions with the pump pulse. This is contrasted by excited state absorption (ESA) and stimulated emission (SE) pathways, where the system is in a single-particle state after the first two interactions with the pump pulse. For ESA pathways the interaction with the probe pulse leads to the creation of a coherence between a two-particle state and the previously excited single-particle state, whereas for SE pathways the interaction with the probe pulse results in a coherence between the GS and a single-particle state. In Fig. \ref{figure2}, we only show the rephasing Liouville space pathways, an equivalent set of diagrams is obtained by interchanging the two interactions with the pump pulse, so-called non-rephasing pathways. However, these two sets of diagrams have the same contribution to the non-linear response, since they only differ in the coherence after the first interaction with the pump pulse, which is not resolved in ordinary PP spectroscopy.

The commutator in the expression of the response function leads to different signs for different Feynman diagrams. The sign of a Feynman diagram is given by $(-1)^n$, where $n$ is the number of interactions with the laser field from the right. While the relative sign between different kinds of Liouville space pathways is fixed by that, there are different conventions for absoulte signs. Here, we will follow the convention that GSB and SE pathways have a negative sign, whereas ESA pathways are positive. 

As an example, how a Liouville space pathway contributes to the response function, the term corresponding to the SE Feynmann diagram $\mathrm{SE_{LP}}$ is given by

\begin{equation}
    S_{\mathrm{SE_{LP}}}^{(3)} (t)=\left(-\frac{i}{\hbar}\right)^3 \mu_{\mathrm{LP}}^4e^{-i \frac{E_{\mathrm{LP}}t}{\hbar}}.
\end{equation}

Between the interaction with the probe pulse and the emission of the non-linear polarization, the off-diagonal element of the density matrix oscillates at the frequency related to the energy difference of the two involved states. We ignore relaxation and dephasing during this time period. All other pathways contribute analogously. The total response function is then obtained by summing over all contributing Liouville space pathways

\begin{equation}
    S^{(3)}(t)=\sum_{i} S_i^{(3)}(t).
\end{equation}

Within the semi-impulsive limit, the laser pulses are assumed to be short compared to any timescale of the system but long compared to the oscillation frequency of the electromagnetic field of the external laser pulse.\cite{mukamel_principles_1995} In this case, the third-order response functions equals the emitted third-order polarization

\begin{equation}
    S^{(3)}(t)=P^{(3)}(t).
\end{equation}

By Fourier transformation of the third-order polarization, one obtains a quantity, whose imaginary part is proportional to the corresponding third-order PP spectrum $\mathrm{PP}^{(3)}(\omega)$ \cite{mukamel_principles_1995, hamm_principles_nodate}

\begin{equation}
    \mathrm{PP}^{(3)}(\omega)\propto-2\mathrm{Im}[P^{(3)}(\omega)].
\end{equation}

Since the Fourier transformation is linear, the PP spectrum can be decomposed into a sum of contributions by the Liouville space pathways. For example, the contribution of the SE Feynman diagram to the PP spectrum is given by

\begin{equation}
    \mathrm{PP}_{\mathrm{SE_{LP}}}^{(3)}(\omega)\propto -\mu_{\mathrm{LP}}^4\delta\left(\omega-\frac{E_{\mathrm{LP}}}{\hbar}\right).
\end{equation}

If homogeneous broadening effects are dominant, the linewidth of the peaks in a PP spectrum is determined by the lifetimes of coherences and populations encountered during the evolution of the systems density matrix.\cite{hamm_concepts_2011} Here, we assume that inhomogenous broadening is dominant, i.e., disorder in the molecular subsystem leads to the observed linewidths. Thus, we replace the delta function in the previous expression by a Gaussian function centered around the same frequency with standard deviation $\sigma = 0.01\ \mathrm{eV}$ for every Liouville space pathway. The resulting peaks are centered around energies corresponding to the frequency at which the off-diagonal elements of the density matrix oscillate during time $t$. 

\section{Results and Discussion}
\subsection{Dependence of PP spectra on the cavity resonance energy}

For the PP spectra of polaritons, there are only six possible transition energies, at which the non-linear polarization is emitted. The dependence of these transition energies as a function of the resonance frequency of the cavity is shown in Fig. \ref{figure3}. Of the six transition energies, three are centered around the energy of the LP while the other three are centered around the energy of the UP.

For our two chosen excitation conditions, either LP or UP excitation with the pump pulse, five different Liouville space pathways contribute to  the non-linear response (Fig. \ref{figure2}). Of each of these five pathways, three emit with approximately the energy of either LP or UP depending on which state was pumped and two emit with approximately the energy of the other respective state. For example, when the LP is pumped, the Liouville space pathways $\mathrm{GSB}^1_{\mathrm{LP}}, \mathrm{SE}^1_{\mathrm{LP}}$ and $\mathrm{ESA}^1_{\mathrm{LP\rightarrow}2\mathrm{LP}}$ contribute to the polarization at approximately the energy of the LP (Fig. \ref{figure3}), while the $\mathrm{ESA}^1_{\mathrm{LP\rightarrow}2\omega}$ and $\mathrm{GSB}^1_{\mathrm{UP}}$ pathways contribute approximately at the energy of the UP. 

The contributions of the pathways at the LP (UP) energies overlap, since the difference at which the polarization is emitted is much smaller than their width. Thus, they cancel each other out for a large part and their strong overlap of positive and negative features leads to the derivative-like lineshape typically observed in PP spectroscopy of polaritons at the LP and UP energies.\cite{delpo_polariton_2020, renken_untargeted_2021, schwartz_polariton_2013} This observation can be rationalized by considering a Holstein-Primakov (HP) transformation \cite{katriel_multiboson_1986} of the TC Hamiltonian. The ensemble of two-level systems in the TC model can be mapped to spin $\frac{1}{2}$ particles. Using the HP transformation, the operators of the ensemble of two-level system can be mapped onto a single pair of bosonic creation and annihilation operators. In this way, the TC Hamiltonian is transformed into a Hamiltonian of two coupled harmonic oscillators.\cite{garraway_dicke_2011}. However, for the TC model this transformation is only exact in the limit of an infinite number of two-level systems, $N\rightarrow\infty$. Since the third order non-linear response of any system of coupled harmonic oscillators vanishes, one expects that also the magnitude of the non-linear response of the TC model decreases as the number of two-level systems is increased. However, for a finite number of two-level systems, as encountered in any real experiment, the non-linear response of the TC model is not exactly zero. From this point of view, the derivative-like spectral features results from an incomplete cancellation of positive and negative signal contributions at the LP and UP energies. Numerically, a decrease in the magnitude of the non-linear response of polaritons with an increasing number of two-level systems has already been observed.\cite{mondal_polariton_2025}

\begin{figure*}
  \includegraphics[]{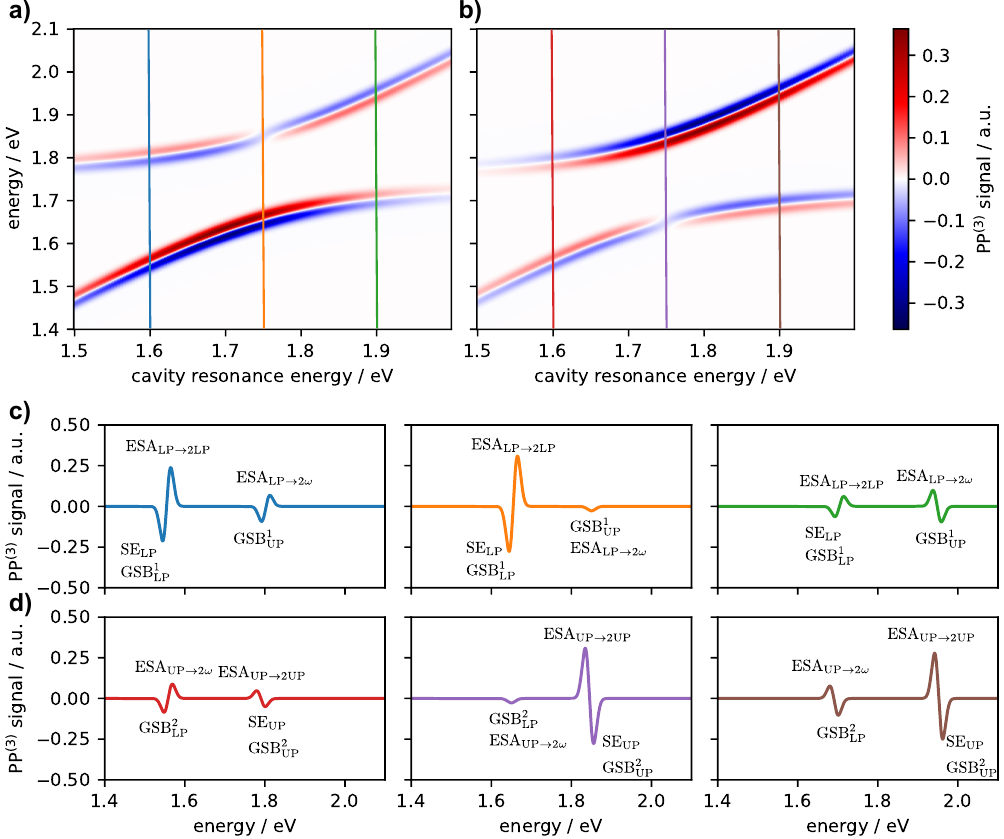}
  \caption{\label{figure4}Lineshapes of pump-probe spectra of polaritons within the TC model with parameters $\hbar\omega_m=1.75\ \mathrm{eV}$, $N=5$, and $\hbar g\sqrt{N}=0.1\ \mathrm{eV}$. a) Lineshape of pump-probe spectra of polaritons when only the LP is excited by the pump pulse depending on the resonance energy of the cavity. b) Lineshape of pump-probe spectra of polaritons when only the UP is excited by the pump pulse depending on the resonance energy of the cavity. c) Cuts along resonance energies of $1.6$, $1.75$, and $1.9\ \mathrm{eV}$ of Fig. \ref{figure4}a. d) Cuts along resonance energies of $1.6$, $1.75$, and $1.9\ \mathrm{eV}$ of Fig. \ref{figure4}b.}
\end{figure*}

For the numerical calculations of the lineshapes we set $N=5$, $\hbar\omega_m=1.75\ \mathrm{eV}$ and $\hbar g\sqrt{N}=0.1\ \mathrm{eV}$. We set the number of coupled two-level system relatively small such that the differences in the transition energies and transition amplitudes are clearly visible (Fig. \ref{figure1} and Fig. \ref{figure3}). Calculations with higher numbers of coupled two-level systems show that our results stay consistent (see SI Fig 1). In Fig. \ref{figure4}a (\ref{figure4}b) the dependence of the lineshape of PP spectra on the cavity resonance energy are shown under LP (UP) pumping conditions. When pumping the LP, the ESA is always blue shifted with respect to the GSB/SE for the spectral feature at the LP energy (Fig. \ref{figure4}a). This is the case, since the transition energy of the LP to 2LP transition is larger than the transition energy of the GS to LP transition for the considered cavity resonance energies (Fig. \ref{figure3}b). For the spectral feature at the energy of the UP however, the ESA is blue shifted with respect to the GSB for cavity resonance energies lower than $1.75\ \mathrm{eV}$ while it is red shifted for resonance energies higher than $1.75\ \mathrm{eV}$ (Fig. \ref{figure4}a). The ESA arises due to the LP to $2\omega$ transition and GSB stems from the GS to UP transition. The transition energies of these two transitions cross at the cavity resonance energy of $1.75\ \mathrm{eV}$ such that the transition energy of the LP to $2\omega$ transition is blue shifted with respect to the GS to UP transition for cavity resonance energies lower than $1.75\ \mathrm{eV}$ and red shifted for cavity resonance energies higher than $1.75\ \mathrm{eV}$. If the cavity is in resonance with the two-level systems, i.e. $\hbar\omega_c=1.75\ \mathrm{eV}$, the derivative-like spectral feature at the UP energy disappears and only a small GSB signal remains. At this resonance energy, the ESA and GSB are emitted with the exact same energy, which leads to a full cancellation of the ESA signal. Since the GS to UP transition amplitude is slightly larger than that of the LP to $2\omega$ transition (Fig. \ref{figure1}c), the contribution of both pathways appear as a single negative peak. 

Similar behavior is observed when pumping only the UP. In this case, the lineshape of the spectral feature at the energy of the UP does not change much with the cavity resonance energy. Here, the ESA is blue shifted with respect to the GSB/SE, since the UP to 2UP transition energy is red shifted with respect to the GS to UP transition energy. However, the lineshape of the spectral feature at the energy of the LP changes as a function of the cavity resonance energy. For cavity resonance energies lower than $1.75\ \mathrm{eV}$, the transition energy of the UP to $2\omega$ transition is larger than that of the GS to LP transition, while this behavior is inverted for energies higher than $1.75\ \mathrm{eV}$. Visually, this leads to a flip of the sign of the derivative-like spectral feature at the LP energy. Similar to before, at exact resonance of the cavity with the two-level systems only a small GSB signal is observed at the energy of the LP.

\subsection{Dark States Relaxation}

Besides the lineshape of the PP spectra of polaritons, where we assumed that no relaxation occurs, we investigate the lineshape of polaritonic systems where LP and UP relax fast to the GS and the DS manifold. To this end, we start from Eq. (14), where the relaxation is introduced via the propagator during the delay time $T$. This means that we assume that all PP spectra are measured at a specific delay time $T=\tau$, at which LP (UP) have completely decayed to the GS and the DS manifold by equal parts, i.e., $50\%$ of the initial population of the LP (UP) have decayed to the GS and $50\%$ of the initial population of the LP (UP) have decayed to the DS manifold. 

\begin{figure*}
  \includegraphics[]{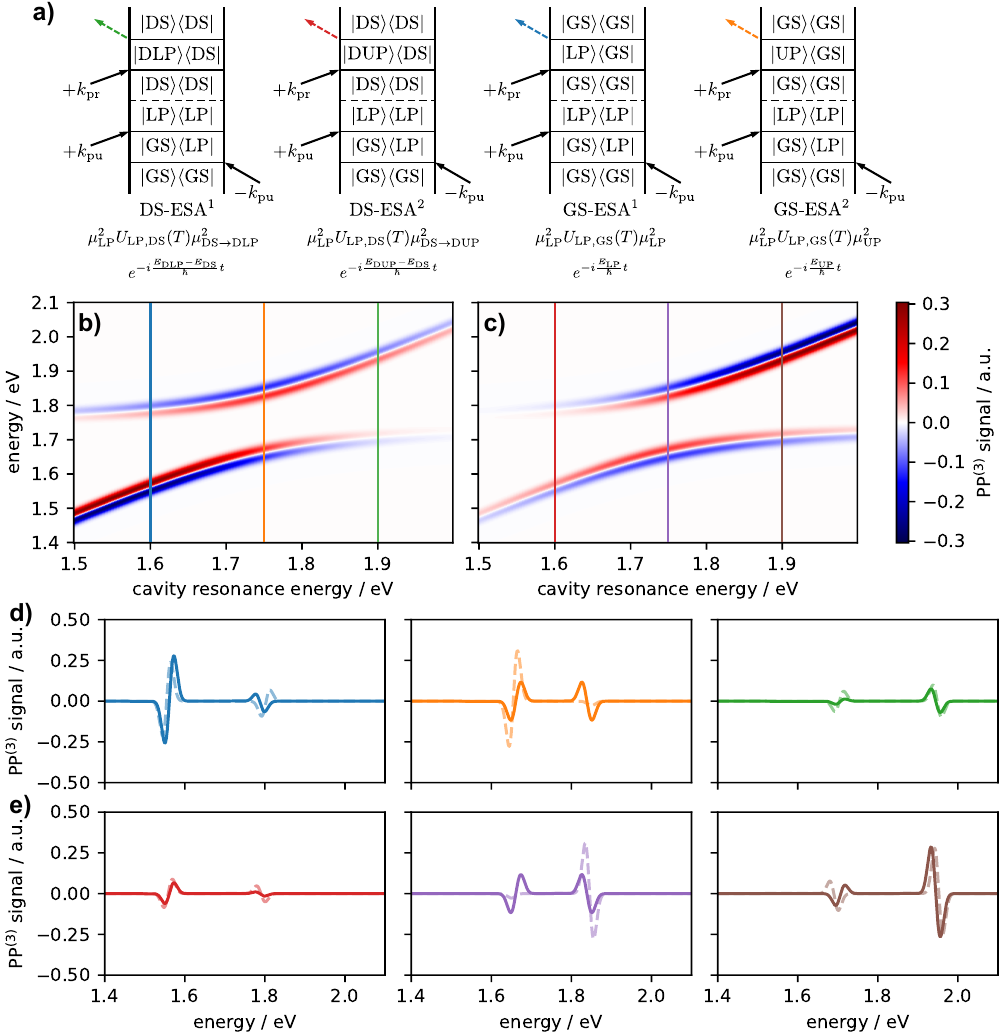}
  \caption{\label{figure5}Dark states relaxation influences the lineshape of pump-probe spectra of polaritons. a) Double-sided Feynman diagrams representing the additional Liouville space pathways, which arise by the relaxation of the LP into the DS manifold and the GS. The color of the last dotted arrow of each Feynman diagrams indicates the energy at which the non-linear polarization is emitted (Fig. \ref{figure3}). b) (c)) Lineshape of pump-probe spectra of polaritons with LP (UP) pumping conditions depending on the resonance energy of the cavity under the assumption that the LP (UP) has completely decayed with equal parts into the DS manifold and the GS. d) (e)) Cuts along resonance energies of $1.6$, $1.75$, and $1.9\ \mathrm{eV}$ of Fig. \ref{figure5}b (\ref{figure5}c). Dotted lines show the cuts without DS relaxation (Fig. \ref{figure4}c).}
\end{figure*}

The DS relaxation leads to various changes in the Liouville space pathways: While all GSB pathways are unaffected by the relaxation of the polaritonic states, SE does no longer contribute to the spectra, since we have assumed that all of the initial population of LP and UP have relaxed either to the GS or the DS manifold. Since the transition moment from any DS to the GS vanishes, there is no SE from the DS to the GS. Furthermore, the ESA pathways are altered such that after the relaxation ESA can only take place via the GS or the DS manifold. ESA starting from the DS is only possible to the DLP or DUP manifolds, since transitions between the DS and polariton-polariton states, i.e., 2LP, 2UP, and $2\omega$, are forbidden. The double-sided Feynman diagrams corresponding to the changed ESA Liouville space pathways if the pump pulse is only resonant with the LP are shown in Fig. \ref{figure5}a. A corresponding set of diagrams for UP pumping is obtain by interchanging LP with UP in the first two interactions with the pump pulse. Relaxation during the delay time $T$ is indicated by a dotted horizontal line in the diagrams of Fig. \ref{figure5}a. While the amplitude of the two respective ESA Liouville space pathways in Fig. \ref{figure2} is diminished by the relaxation, the four shown Liouville space pathways in Fig. \ref{figure5}a become non-negligible. For example, the contribution of the $\mathrm{DS\text{-}ESA^1}$ pathway to the third-order PP spectrum is given by

\begin{equation}
    \mathrm{PP}_{\mathrm{DS\text{-}ESA^1}}^{(3)}(\omega)\propto \mu_{\mathrm{LP}}^2U_\mathrm{LP,DS}(T)\mu_{\mathrm{DS\rightarrow DLP}}^2\delta\left(\omega-\frac{E_{\mathrm{DLP}}-E_{\mathrm{DS}}}{\hbar}\right),
\end{equation}

where $U_\mathrm{LP,DS}(T)$ is the matrix element of the propagator, which equals the probability, that given that the system is in the LP state at $T=0$, it is in a DS at $T=\tau$. Under our assumptions, we have $U_\mathrm{LP,DS}(\tau) = 0.5$. We have also carried out calculations with different ratios of GS and DS population after the relaxation. These calculations show that only the amplitude of the signal is effected but not the shape of the lineshapes themselves (see SI Fig. 2).

The dependence of the lineshape of the third-order PP spectra as a function of the cavity resonance energy for LP (UP) pumping are shown in Fig. \ref{figure5}b (\ref{figure5}c). As in the case without relaxation, the third-order PP spectra consist of derivative-like spectral features at the LP and UP energies. After the relaxation, ESA only takes place via the DS to DUP and DLP transitions and negative signals result from the GSB at LP und UP energies. Therefore, the derivative-like spectral features emerge, because the transition energies of the DS to DLP (DUP) transitions differ from those of the GS to LP (UP) transitions (Fig. \ref{figure3}b and \ref{figure3}d). The resulting lineshape at the energy of the pumped state looks similar to the lineshape at the energy of the pumped state without relaxation. However, at the energy of the state, which was not pumped, the ESA signal is always red (blue) shifted with respect to the GSB signal for LP (UP) pumping, since the transition energy for the DS to DUP (DLP) transition is always red (blue) shifted compared to the GS to UP (LP) transitions. Whereas without relaxation, the derivative-like feature at the UP (LP) energy is inverted when the cavity is tuned through the resonance of the two-level systems under LP (UP) pumping conditions.

Here, we only show the dependence of the lineshape in PP spectra on the cavity resonance energy. However experimentally, it is challenging and time costly to measure the dependence of the lineshape on the cavity resonance energy. It necessitates an accurate adjustment of the incidence angle of the laser pulses onto the sample such that one is able to precisely address modes with different cavity resonance energies.\cite{rodel_anisotropic_2025, buttner_probing_2025} Furthermore, the pump laser might have to be continuously tuned to the energy of LP or UP, whose energies change when the cavity resonance energy is varied.

An easier way to reveal the relaxation from the polariton states to the DS is by performing time-dependent measurements at a fixed incidence angle. At short delay times, before DS relaxation takes place, one would observe a lineshape according to Fig. \ref{figure4}. This lineshape then transforms into that shown in Fig. \ref{figure5} if the delay time is continuously increased. Hereby, the incidence angle has to be chosen such that the cavity resonance energy is red (blue) shifted with respect to the molecular transition energy if the LP (UP) is pumped. This ensures that a flip in the sign of the derivative-like feature at the energy of the state, which was not pumped, occurs. Generally, such experiments necessitate ultra short laser pulses, whose temporal widths are smaller than the time constant of the relaxation process from the polariton states to the DS. This time constant has to be on the same order as the time constant of the relaxation process from the polariton states to the ground state such that non-negligible population is transferred to the DS. The time constant of the relaxation from the polariton states to the DS is mainly determined by the lifetime of the cavity photons, which highly depends on the type of cavity. In microcavities, the cavity photons usually life for a few tens of femtoseconds, while the lifetime of surface plasmon polaritons is around 100 fs.\cite{fassioli_femtosecond_2021, buttner_probing_2025} If the temporal width of the laser pulses is not chosen accordingly, the changes in the lineshape cannot be temporally resolved.

\section{Conclusion}

We have combined a quantum mechanical model for the description of polaritons, the TC model, with the response function formalism to simulate PP spectra under electronic strong-coupling conditions. First, we used our model to investigate the dependence of the lineshape in PP spectra on the resonance energy of the cavity, which experimentally corresponds to the dependence on the incidence angle of the laser pulse onto the polaritonic sample. The simulated spectra show derivative-like lineshapes at the LP and UP energies as observed in many experiments.\cite{delpo_polariton_2020, renken_untargeted_2021, schwartz_polariton_2013} The treatment within the response function formalism, allowed us to assign photo-physical processes to the observed signals in PP spectra, i.e., GSB, SE and ESA of the polariton states. The dependence of the lineshape in the PP spectra on the resonance energy of the cavity enables a clear distinction of the response of polaritons from that of possible untargeted effects.

Subsequently, we phenomenologically introduced a fast relaxation of the polariton states into the GS and the DS manifold. By this, we were able to simulate the response of the system after relaxation and to investigate the lineshape in PP spectra again depending on the resonance energy of the cavity. The comparison of the lineshapes with and without relaxation revealed that the relaxation to the DS manifold results in a change of the lineshape, i.e., a flip in the sign of the derivative-like feature at the energy of the state, which was not pumped. Moreover, we argue that this change in the lineshape induced by the DS relaxation is observable by performing time-dependent measurements at a fixed incidence angle. In this case, the DS relaxation leads to a time-dependent change in the lineshape, i.e., a flip of the sign of the derivative-like spectral feature at the energy of the UP (LP) if the LP (UP) is initially excited by the pump pulse and the incidence angle is chosen such that the cavity resonance energy is red (blue) shifted with respect to the molecular transition energy.

Up to our knowledge, there has been no experimental observation of a change in the lineshape in time-dependent PP spectra of polaritons induced by relaxation of the polariton states into the DS manifold so far. We think that this is due to the fact that most PP experiments of polaritonic systems are conducted with laser pulses with temporal widths on the same order of magnitude as the lifetime of the cavity photons.\cite{schwartz_polariton_2013, delpo_polariton_2020, renken_untargeted_2021, rozenman_long-range_2018, avramenko_quantum_2020, takahashi_singlet_2019} Since theoretical studies found that the relaxation to the DS manifold occurs on the same time scale as the cavity photons decay, changes in the lineshape cannot be resolved using laser pulses with the same temporal length as the lifetime of the cavity photons.\cite{sokolovskii_non-hermitian_2024, tichauer_multi-scale_2021}

Our results enable the detailed analysis and interpretation of PP spectra of polaritons and allow experimentalists to decide whether they observe the response of polaritons or other untargeted effects. Understanding and being able to track the photoinduced dynamics of polaritonic systems in the electronic strong-coupling regime is a first step towards their application for optoelectronic devices.

\section*{Supplementary Material}
\vspace*{-2ex}
See supplementary material for details on Liouville space pathways for UP pumping with relaxation, $N$-dependence of PP lineshapes, dependence of the PP lineshapes on different relaxation ratios, PP lineshapes with disorder, and PP lineshapes with transition dipole moment.

\begin{acknowledgments}
 L.N.P. acknowledges a fellowship by the FCI. J.L. acknowledges support from the HFSP fellowship program under Grant No. LT0056/2024-C.
\end{acknowledgments}

\section*{Conflicts of interest}
\vspace*{-2ex}
There are no conflicts to declare.

\section*{Data availability}
\vspace*{-2ex}
The data that support the findings of this study are available from the corresponding author upon reasonable request.

\bibliography{Polariton_lineshapes}

\end{document}

%\preprint{}

\title{Supplementary Material: Lineshapes in Pump-Probe Spectroscopy of Polaritons} %Title of paper



\author{Luca Nils Philipp}
\affiliation{Institut für Physikalische und Theoretische Chemie, Universität Würzburg, Emil-Fischer Straße 42, 97074 Würzburg, Germany}
\author{Eva Münzel}
\affiliation{Institut für Physikalische und Theoretische Chemie, Universität Würzburg, Emil-Fischer Straße 42, 97074 Würzburg, Germany}
\author{Julian Lüttig}
\affiliation{Department of Physics, University of Michigan, 450 Church Street, Ann Arbor, Michigan 48109, USA}
\affiliation{Institut für Physikalische und Theoretische Chemie, Universität Würzburg, Am Hubland, 97074 Würzburg, Germany}
\author{Roland Mitri\'c}
\email{roland.mitric@uni-wuerzburg.de}
\affiliation{Institut für Physikalische und Theoretische Chemie, Universität Würzburg, Emil-Fischer Straße 42, 97074 Würzburg, Germany}


\date{\today}

\pacs{}% insert suggested PACS numbers in braces on next line

\maketitle %\maketitle must follow title, authors, abstract and \pacs

\section{Liouville space pathways for UP pumping with relaxation}

\begin{figure*}[!ht]
\includegraphics[]{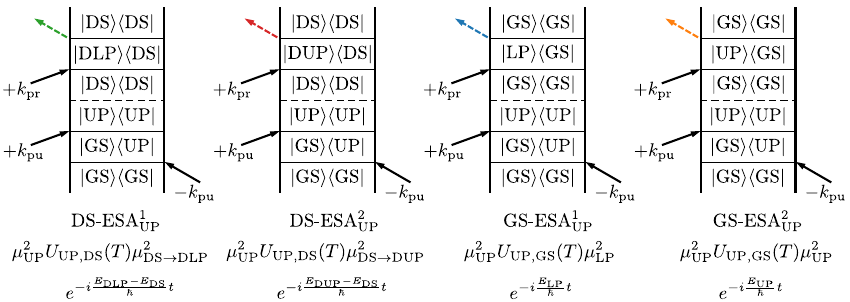}
\caption{Double-sided Feynman diagrams representing the additional Liouville space pathways, which arise by the relaxation of the UP into the DS manifold and the GS. The color of the
last dotted arrow of each Feynman diagrams indicates the energy at which the non-linear
polarization is emitted (Fig. 3).}
\label{S1}
\end{figure*}

\newpage

\section{$N$-dependence of the PP lineshapes}

\begin{figure*}[ht]
  \includegraphics[]{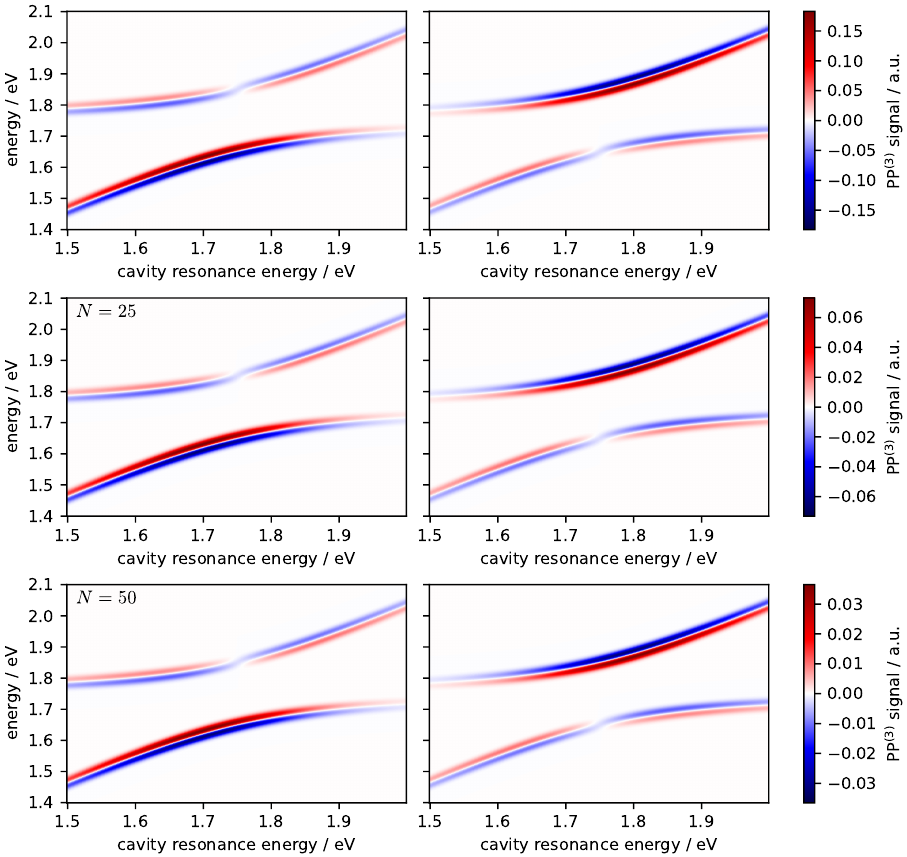}
  \caption{Lineshape of pump-probe spectra
of polaritons when LP (left column) or UP (right column) is excited by the pump pulse depending on the resonance
energy of the cavity for different numbers of two-level systems in the Tavis-Cummings model.}
\label{S2}
\end{figure*}

\newpage

\section{Dependence of the PP lineshapes on different relaxation ratios}

\begin{figure*}[ht]
  \includegraphics[]{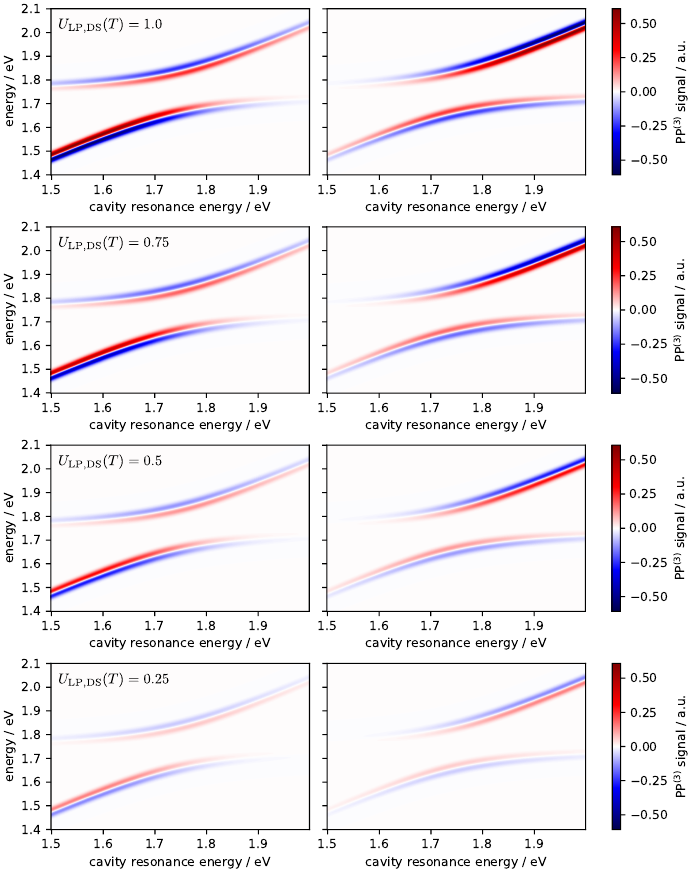}
  \caption{Lineshape of pump-probe spectra
of polaritons when LP (left column) or UP (right column) is excited by the pump pulse depending on the resonance
energy of the cavity after relaxation of the polaritonic states into the DS and GS for different relaxation rate ratios between polariton to DS and polariton to GS processes.}
\label{S3}
\end{figure*}

\newpage

\section{PP Lineshapes with onsite disorder}

\begin{figure*}[ht]
  \includegraphics[]{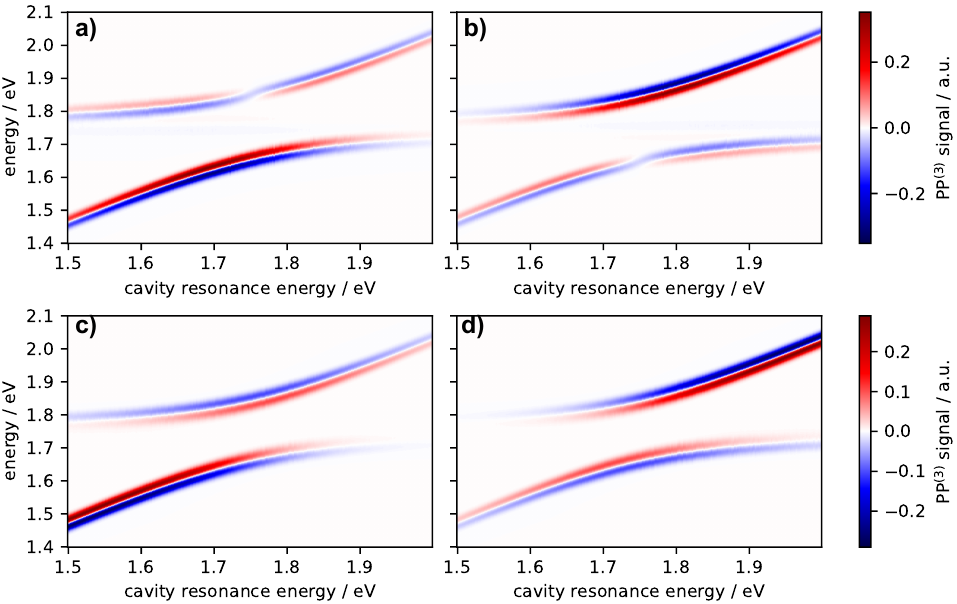}
  \caption{a) (b)) a) Lineshape of pump-probe spectra of polaritons when only the LP (UP) is excited by the pump pulse depending on the resonance energy of the cavity. c) (d)) Lineshape of pump-probe spectra of polaritons with LP (UP) pumping conditions depending on the resonance energy of the cavity under the assumption that the LP (UP) has completely decayed with equal parts into the DS manifold and the GS.}
  \label{S5}
\end{figure*}

\begin{figure*}[]
  \includegraphics[]{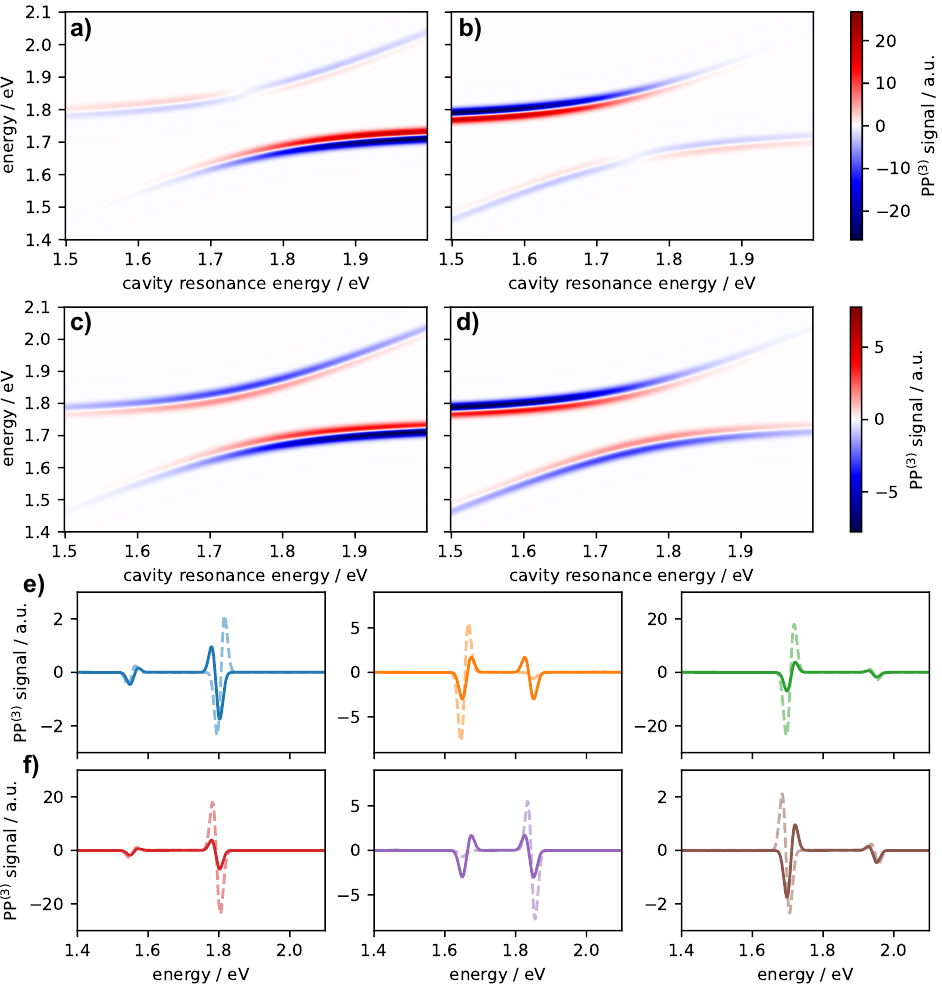}
  \caption{a) (b)) a) Lineshape of pump-probe spectra of polaritons when only the LP (UP) is excited by the pump pulse depending on the resonance energy of the cavity. c) (d)) Lineshape of pump-probe spectra of polaritons with LP (UP) pumping conditions depending on the resonance energy of the cavity under the assumption that the LP (UP) has completely decayed with equal parts into the DS manifold and the GS. d) (e)) Cuts along resonance energies of $1.6$, $1.75$, and $1.9\ \mathrm{eV}$ of Fig. S\ref{S4}c (S\ref{S4}d). Dotted lines show the cuts without DS relaxation.}
  \label{S4}
\end{figure*}

To include onsite disorder in our model, we randomly add small frequency deviations to the resonance frequency of the two-level systems in the Tavis-Cummings Hamiltonian. In this case, the Hamiltonian is given by

\begin{equation}
    H_\mathrm{TC} = \hbar\sum_{i=1}^N (\omega_m+\delta_i)\sigma_i^+\sigma_i^- + \hbar\omega_ca^\dag a + \hbar g \sum_{i=1}^N\left(a^\dag\sigma_i^- + \sigma_i^+a\right),
\end{equation}
 
where $\delta_i$ is a small frequency deviation sampled from a uniform distribution on $[-\frac{\Delta}{\hbar}, \frac{\Delta}{\hbar}]$. We have used the Tavis-Cummings Hamiltonian with onsite disorder to simulate the dependence of the lineshape of PP spectra on the resonance energy of the cavity with and without relaxation. To this end, we use the parameters as in the main manuscript, set $\Delta = 0.05\ \mathrm{eV}$ and averaged over 100 disorder realizations. The resulting lineshapes are shown in Fig. S\ref{S4}. As evident from Fig. S\ref{S4}, onsite disorder has very little influence on the dependence of the lineshape of PP spectra on the cavity resonance energy. Furthermore, by including disorder the ground state  to dark states transition acquires very small transition moments, which leads to the faint nonlinear signals at an energy of $1.75\ \mathrm{eV}$. Similar behavior has been observed in several other studies.\cite{reitz_nonlinear_2025, schwennicke_extracting_2024, sommer_molecular_2021}

\section{PP Lineshapes with transition dipole moment}

\newpage

\bibliography{Polariton_lineshapes}